\documentclass[12pt]{article}
\usepackage{graphics}
\begin{document}

\thispagestyle{empty}

\begin{center}
{\large\bf Local Bursts Model of CMB Temperature Fluctuations:
Scattering in Primordial Hydrogen Lines}
\end{center}

\begin{center}
V. K. Dubrovich$^1$, S. I. Grachev$^2$
\end{center}

\begin{center}
{\small
$^1$Special Astrophysical Observatory, St. Petersburg Branch, Russian
Academy of Sciences, St. Petersburg, 196140 Russia\\
$^2$Sobolev Astronomical Institute, St. Petersburg State University,
Universitetskii pr. 28, St. Petersburg, 198504 Russia\\
{\it Journal ref.: Astronomy Letters, 2015, Vol. 41, No. 10, pp. 537-548}}
\end{center}

{\small

Within the framework of a flat cosmological model a 
propagation of an instantaneous burst of isotropic
radiation is considered from the moment of its beginning at some initial
redshift $z_0$ to the moment of its registration now (at $z=0$). Thomson 
scattering by free electrons and scattering in primordial hydrogen lines
H$_\alpha$, H$_\beta$, P$_\alpha$ and P$_\beta$ are considered as the sources of
opacity and when calculating the albedo of single scattering in the lines we
take into account deactivation of the upper levels of transitions by
background blackbody radiation. Profiles for these lines in a burst spectrum
are calculated for
different distances from the center of the burst and different values of $z_0$.
In a first approximation these profiles do not depend on spectrum and intensity
of a burst radiation.
It is shown that lines are purely in absorption at sufficiently large
distance but
emission component may appear as a distance decreases and it becomes stronger 
while absorption component weakens with a further distance decrease.
For the sum of H$_\alpha$ and H$_\beta$ lines the depth of absorption can
reach $2\cdot 10^{-4}$ while for the sum of P$_\alpha$ and P$_\beta$ lines the
maximum absorption is about $7\cdot 10^{-6}$. So that the relative magnitude
of temperature fluctuations lies between 10$^{-7}$ and 10$^{-9}$. The
calculations were fulfiled for bursts with different initial sizes. For the
same $z_0$ the profiles of hydrogen lines are practically coinside for burst 
sizes lower than someone and for greater ones the lines weaken as the burst
size grows.
\bigskip


\medskip

}

\newpage

\begin{center}
{\bf Introduction}
\end{center}

Investigations of cosmic microwave background (CMB) still continue. New
results obtained by PLANCK mission (see Adam et al., 2015a) define our
present day knowledge about power spectrum of primordial spatial fluctuations
of matter density, about some global fundamental parameters of the Universe
and about CMB polarization. Essencial progress is achieved also in CMB
spectroscopy. However all these impotant successes do not exclude further more
detailed and deep investigations.

In particular medium and high resolution spectroscopy of separate (indvidual)
objects (elements of CMB brightness map) seems to be very important. Novelty
here is in turning from investigation of statistical CMB properties
defined by global processes in the early Universe to searching and learning
local phenomena and objects. The last ones can be somewhat rare events not
affecting on the average statistical parameters of CMB. But they can carry
information about physical laws. So for example one can expect new local forms
of matter and fields (see, e.g. Dubrovich (2003), Grachev and Dubrovich (2011),
Dubrovich and Glazyrin (2012)).

Besides of more or less probable but still hypothetical objects there exists
evidently the whole class of local sources in the early Universe which can be
learned individually. These are the same standard primordial CMB temperature
fluctuations thoroughly learned now statistically. In fact we deal with some
spatial domains of space where temprature increase or decrease takes place
for some reason or other. It is very important that besides spatial apartness
of these regions their temperature deviations are also nonstationary.
Depending on a mechanism of a given inhomogeneity formation a typical time of
its development and damping can be different. So for example if CMB temperature
fluctuation forms due to acoustic waves then a typical time $\Delta t$ of its
life will be of the order of a wave oscillation period i.e. for a
fluctuation scale  $L$ and sound speed $c_{\rm s}=c/\sqrt{3}$ ($c$ is
the speed of light) will be $\Delta t\approx L/c_{\rm s}$ or
$\Delta t/t_0\approx \vartheta{\sqrt{z_0}}$ where $t_0$ is a cosmological
time at the moment of fluctuation appearance and $\vartheta$ is its
present-day angular size. For the angular size $\vartheta\approx 3'$ we have
$\Delta t/t_0\approx 3.5\cdot 10^{-2}$. This time of life can be considered
as a burst. On the other hand after approximately the same time the sign of
the effect in this region changes and after that repeats oneself again. Then
one needs to count up a summary effect. If however there is an accidental
interference of several acoustic waves with different wave vectors in the
same region of space then the resaltant fluctuation may have appreciably larger
amplitude but lower probability of a recurrance of such an event in a given
spatial domain. Real observations give maximum amplitude of temperature
deviation in the spots of the order of 500 $\mu$K (Adam et al., 2015b). So we
have in fact local burst model of sources in the early Universe within the
framework of standard scenario of matter evolution without any additional exotic
hypothesies. Observed now CMB temperature map is a sum of all sources with
account of their initial spatial and temporal distribution and with account
of subsequent scattering of photons emitted by them. These two factors must
be taken in a product of one on another. In this work we calculate a function
of transition from radiation intensity of these sources to observed intensity
taking into account scattering on free electrons and in hydrogen subordinate
lines in a space between a source and an observer. In a first approximation
this function does not depend on a source intensity.

Existing theoretical formulae correctly take into account all these effects
for Thomson scattering on free electrons. However in the case of scattering
in spectral lines some new additional effects take place. The most important
(end evident) distinction is a different dependence on frequency: Thomson
scattering is the same for all photons in a given place and in a given time
irrespective of their frequencies but scattering in lines involves only
photons in a very narrow frequency band. In expanding Universe every spatial
point and every moment of time define a redshift (observed frequency in fact)
of photons emitted in this point and at this moment of time. Summing along a
line of sight contributions from different spatial domains with a proper
account of photons emission time in the case of Thomson scattering leads to
averaging of fluctuations with different signs. In the case of scattering in
lines CMB background distortions formed by different layers will be seen now
on different frequencies. So frequency becomes the third coordinate which can
be used to analyse the physics and parameters of evolution processes.

Important factor is an optical thickness due to scattering. Scattering on free
electrons is multiple one because the smallness of its cross-section is
compensated by comparatively large photon path length and electron 
concentration. In hydrogen subordinate lines we have small optical thickness
because scattering takes place on a small piece of path (defined by a width of
line profile) and occupation numbers of excited atomic levels are comparatively
small. It means that for scattering in lines we may confine ourselves to a
single scattering approximation.

A various contribution of nonconservative effects of photons scattering in a
resultant spectrum is a more fine and not at all evident distinction. In
particular we mean deactivation of excited levels due to absorption of
background photons or due to collisions with electrons and atoms. These effects
are taken into account by introducing in equations of radiative transfer
so-called albedo of a single scattering $\lambda$. Detailed account of
nonconservative scattering effects is necessary because in our case transfer of
radiation from these sources is nonequilibrium: due to small optical
thicknesses in subordinate lines both occupation numbers of levels and
radiation spectrum do not have a time to be thermalized.

In our case albedo of a single scattering in hydrogen lines is determined by
photoionization and radiative transitions due to absorption of CMB photons.
Simple estimates show that in this case $\lambda$ can be noticably
lower than unity in contradiction to scattering on free electrons with an
account of double Compton effect when $\lambda=1$ practically. As will be
shown below this distinction is very important and it has significant
influence on a resultant spectrum.

The present paper is a continuation of our previous work (Grachev and
Dubrovich, 2011) devoted to calculation of a radiation field evolution of an
instantaneous burst of isotropic radiation as a result of Thomson scattering
in a homogeneous expanding and recombining Universe. Now along with Thomson
scattering we take into account scattering in hydrogen subordinate lines
H$_\alpha$, H$_\beta$, P$_\alpha$ and P$_\beta$ and we calculate line profiles
in a burst spectrum on different distances from the burst center for
different burst moments  $z_0$. We do not take into account radiation
polarization and we consider scattering (both Thomson one and in lines) as
isotropic. Moreover we consider scattering in lines to occur with complete
frequency redistribution (CFR), so that source function do not depend on
frequency. We also assume that the burst radiation has no influence on electron
concentration and on occupation numbers of atomic energy levels which are
calculated beforehand using our code of primordial hydrogen recombination
dynamics (Grachev and Dubrovich, 1991).

Rubi\~{n}o-Martin et al. (2005) learned an influence of scattering in
subordinate lines of primordial hydrogen on the theoretical power spectrum
of CMB angular fluctuations. But we calculate profiles of spectral lines
arising as a result of scattering of radiation from an external (with respect
to CMB) source. Moreover as it was noticed above we take into
account nonconservative character of radiative transfer in lines by means of
albedo of a single scattering which is completely absent in previous papers.
\begin{center}
{\bf Main equations and relations}
\end{center}

We consider transfer of radiation in a homogeneous expanding Universe for
initial spherically-symmetrical distribution of radiation intensity. For a flat
model of the Universe corresponding nonstationary scalar transfer equation for
photons occupation number $n$ is (Nagirner and Kirusheva, 2005):
\begin{equation}
\frac{\partial n}{\partial \eta}+\mu\frac{\partial n}{\partial r}
+\frac{1-\mu^2}{r}\frac{\partial n}{\partial \mu}-\frac{a\,\nu}{c}H
\frac{\partial n}{\partial\nu}=-a\,k_{\rm e}(n-s_e)-a\,k_{ik}(n-s_{ik}),
\label{eq1}
\end{equation}
where $\eta$ is a conformal time ($d\eta=cdt/a(\eta)$, $c$ is the speed of
light, $t$ is a time, $a=a(\eta)$ is the cosmologicalscale factor), $r$ is
a distance parameter, $H=H(\eta)$ is the Hubble factor, $n=(c^2/2h\nu^3)I$,
$I=I(r,\mu,\eta, \nu)$ is a radiation intensity at a frequency $\nu$
propagating at angle $\vartheta=\arccos\mu$ to a radial direction, $s_e(r,\eta,
\nu)$ and $s_{ik}(r,\eta)$ are dimensionless (in average occupation numbers) 
source functions for Thomson scattering and for scattering in a spectral line 
for a transition between lower energy level $i$ and upper energy level $k$.
Volume extiction coefficients corresponding to these two types of scattering are
\begin{equation}
k_e(\eta)=\sigma_e n_e(\eta),\quad k_{ik}(\eta,\nu)=\overline{k}_{ik}(\eta)
\varphi_{ik}(\nu),\quad\overline{k}_{ik}(\eta)=\frac{h\nu_{ik}}{4\pi}n_i
B_{ik}\left(1-\frac{n_kg_i}{n_ig_k}\right),
\label{kekik}
\end{equation}
where $\sigma_e=6.65\cdot 10^{-25}$ cm$^2$ is the cross-section of Thomson
scattering, $n_e(\eta)$ is an electron concentration, $h\nu_{ik}$ is an energy
of transition, $B_{ik}$ is the Einstein coefficient for absorption of radiation,
$n_i(\eta)$ and $n_k(\eta)$ are occupation numbers of levels and $g_i$ and
$g_k$ are their statistical wheights, $\varphi_{ik}(\nu)$ is an absorption
coefficient profile normalized as $\int_0^\infty \varphi_{ik}(\nu)d\nu=1$.
Note that in a volume coefficient of extinction in a line we consider induced
radiation as a negative absorption.

We consider scattering both on electrons and in lines to be isotropic and
scattering in lines to be completely incohernt as well. Therefore for radiation
field with the axial symmetry the source functions in (\ref{eq1}) are written as
\begin{equation}
s_e(r,\eta,\nu)=j(r,\eta,\nu), \quad s_{ik}(r,\eta)=\lambda_{ik}(\eta)
j_{ik}(r,\eta),
\label{Sfun}
\end{equation}
where $\lambda_{ik}(\eta)$ is an albedo of a single scattering in a line,
\begin{equation}
j(r,\eta,\nu)=(1/2)\int_{-1}^{+1}n(r,\mu,\eta,\nu)d\mu, \quad 
j_{ik}(r,\eta)=\int_0^\infty j(r,\eta,\nu)\varphi_{ik}(\nu)d\nu
\label{jjik}
\end{equation}
are radiation intesities averaged over directions and both over directions and
over an absorption coefficient profile. When calculating a single scattering
albedo we take into account spontaneous transitions and also transitions
induced by blackbody CMB radiation with the exception of transitions in
Lyman lines being optically very thick. As a result we have
\begin{equation}
\lambda_{ik}(\eta)=R_{ki}/[\sum_{i'=2,i'\neq k}^{\infty}R_{ki'}(\eta)+
R_{kc}(\eta)],\quad k>i,
\label{lik}
\end{equation}
where
\begin{equation}
R_{ki}(\eta)=\frac{g_i}{g_k}\frac{A_{ik}}{\exp[h\nu_{ki}/kT(\eta)]-1},\quad k<i,
\label{Rkid}
\end{equation}
are the coefficients of transition probabilities upwards due to absorption
of the blackbody radiation with the temperature $T(\eta)$ and
\begin{equation}
R_{ki}(\eta)=\frac{A_{ki}}{1-\exp[-h\nu_{ik}/kT(\eta)]},\quad k>i,
\label{Rkiu}
\end{equation}
are the coefficients of transition probabilities for spontaneous and induced
(due to the blackbody radiation) transitions downwards, $R_{kc}$ are
the coefficients of transition probabilities due to ionization by the
blackbody radiation, $A_{ki}$ are the Einstein coefficients for spontaneous
transitions.

Further we neglect both natural and Doppler (due to thermal motion of atoms)
widths of lines as compared with the width defined by cosmological space
expansion. So we consider the line profile as the delta-function in a
comoving frame of reference: $\varphi_{ik}(\nu)=\delta(\nu-\nu_{ik})$ where
$\nu_{ik}$ is the frequency of transition in the laboratory frame. Then
according to eqs. (\ref{Sfun}) and (\ref{jjik})
\begin{equation}
s_{ik}(r,\eta)=\lambda_{ik}(\eta)j(r,\eta,\nu_{ik}),
\label{jikd}
\end{equation}
and one can rewrite the wrighthand side of eq. (\ref{eq1}) to obtain as a
result the following basic equation:
\begin{equation}
\frac{\partial n}{\partial \eta}+\mu\frac{\partial n}{\partial r}
+\frac{1-\mu^2}{r}\frac{\partial n}{\partial \mu}-\frac{a\,\nu}{c}H
\frac{\partial n}{\partial\nu}=-a\,k_{\rm e}(n-s)-a\,\overline{k}_{ik}\delta
(\nu-\nu_{ik})(n-\lambda_{ik}s),
\label{maineq}
\end{equation}
where the source function $s=s_e=j(r,\eta,\nu)$.

Hereinafter the space expansion factor $a$ is considered to be 1 in the
present epoch ($z=0$) so that $a(z)=1/(1+z)$ where $z$ is a redshift.
Therefore the distance coordinate $r$ in eq. (\ref{maineq}) is the distance
from the center of symmetry measured in the present time (at $z=0$).
For an arbitrary $z$ the corresponding distance is $r\,a(z)=r/(1+z)$.

The problem is solved by an ordinary method widely used in the radiative
transfer theory. At first one gets analytical formal (i.e. for a given
source function $s(r,\eta,\nu)$) solution of the main eq. (\ref{maineq})
for a given initial distribution of radiation intensity. Then this solution
is substituted into the definition of the source function through the
radiation intensity and one gets an integral equation for the source function
which does not depend on direction. This equation is solved by any one
numerical method and then the radiation field with its angular structure
is obtained from the formal solution by means of simple numerical integration.
We have used this method in our preceding paper. The complication consists in
in the presence of frequency dependence of the main functions. However since
Thomson scattering is neutral (i.e. its cross-section does not depend on
a frequency) and moreover we neglect frequency change for line scattering in
the comoving frame then frequency changes due to cosmological space expansion
only: $\nu a(\eta)=\nu' a(\eta')$ or $\nu'=\nu (1+z')/(1+z)$ where $z$ is
the redshift corresponding to the moment $\eta$. Therefore one should take into
account cosmological change of frequency as well when deriving the formal
solution.

In order to obtain the formal solution mentioned above let us consider
propagation of radiation along a ray intersecting radial direction at an
angle $\vartheta=\arccos\mu$ on a distance $r$ from the center of symmetry.
Let $l$ be a coordinate measured from the nearest to the center of symmetry
point of the ray. The lefthand side of eq. (\ref{maineq}) represents
directional derivative along the ray and so this equation assumes the form
\begin{equation}
dn/dl=-\alpha(l,\nu(l))n(l,\nu(l))+\beta(l,\nu(l))s(l,\nu(l)),
\label{meql}
\end{equation}
where
\begin{equation}
\alpha(l,\nu(l))=a(l)k_e(l)+a(l)\overline{k}_{ik}(l)\delta(\nu(l)-\nu_{ik}),
\end{equation}
\begin{equation}
\beta(l,\nu(l))=a(l)k_e(l)+a(l)\overline{k}_{ik}(l)\delta(\nu(l)-\nu_{ik})
\lambda_{ik}(l).
\end{equation}
Integrating this equation we obtain
\begin{equation}
n(l,\nu(l))=n_0(l_0,\nu(l_0))e^{-\int_{l_0}^l\alpha(l',\nu(l'))dl'}+
\int_{l_0}^l\beta(l',\nu(l'))s(l',\nu(l'))e^{-\int_{l'}^l\alpha(l'',\nu(l''))
dl''}dl'.
\label{fsoll}
\end{equation}
Here $l_0$ is the coordinate of the point farthest from the observation point
$l$ but yet capable to give a contribution to radiation field in the point $l$
at the moment of time $\eta$. Every point $l'$ on the ray is defined by the
radial distance $r'$ an by the angle $\arccos\mu'$ between the ray and radial 
direction. From geometry of the problem simple relations follow:
\begin{equation}
l=r\mu,\quad l'=r'\mu',\quad l_0=r_0\mu_0,\quad r\sqrt{1-\mu^2}=
r'\sqrt{1-\mu'^2}
\label{rrll}
\end{equation}
and
\begin{equation}
l-l_0=\eta,\quad l-l'=\eta-\eta'.
\label{ll}
\end{equation}
According to the last of these equations one can turn to integration over time
in eq. (\ref{fsoll}) since $dl'=d\eta'$. As a result taking into account eqs.
(\ref{rrll}) and (\ref{ll}) the formal solution (\ref{fsoll}) is written in
the following form
\begin{equation}
n(r,\mu,\eta,\nu)=n_0(r_0,\mu_0,\nu_0)e^{-\int_0^\eta\alpha(\eta'',\nu'')
d\eta''}+\int_0^\eta\beta(\eta',\nu')s(r',\eta',\nu')e^{-\int_{\eta'}^\eta
\alpha(\eta'',\nu'')d\eta''}d\eta',
\label{fsol}
\end{equation}
where $n_0(r_0,\mu_0,\nu_0)$ is initial (at the moment $\eta=0$) radiation
distribution over distances from the burst center and over angles and
frequencies. Here
\begin{equation}
\nu_0=
\nu a(\eta)/a(0),\quad \nu'=\nu a(\eta)/a(\eta'),\quad  \nu''=
\nu a(\eta)/a(\eta''),
\label{nunu}
\end{equation}
\begin{equation}
r_0=\sqrt{r^2-2r\mu\eta+\eta^2},\quad r_0\mu_0=r\mu-\eta,
\label{rhomu0}
\end{equation}
\begin{equation}
r'=\sqrt{r^2-2r\mu(\eta-\eta')+(\eta-\eta')^2},\quad r'\mu'=
r\mu-\eta+\eta'.
\label{rhomup}
\end{equation}

Instead of conformal time measured in the length units one can introduce
dimensionless time
\begin{equation}
u=\int_0^\eta a(\eta')k_e(\eta')d\eta'=c\sigma_e\int_0^t n_e(t')dt'=
c\sigma_e\int_z^{z_0} \frac{n_e(z')}{(1+z')H(z')}dz,'
\label{udef}
\end{equation}
which has a physical sence of optical path length (for Thomson scattering)
between the moments $z$ and $z_0$. Here redshift $z_0$ corresponds to
initial moment of time i.e. $u=\eta=t=0$ at $z=z_0$. Conformal time $\eta$
and dimensionless time $u$ can be obtained by means of relation
\begin{equation}
\eta=c\int_0^{t}dt'/a(t')=c\int_z^{z_0}dz'/H(z'),
\label{etaz}
\end{equation}
when calculating $u$ and $\eta$ on the same redshift grid. Here $H(z)$ is the
Hubble factor. Let us consider the integral in eq. (\ref{fsol}):
\begin{equation}
\int_0^\eta \alpha(\eta'')d\eta''=u+\int_0^\eta a(\eta'')\overline{k}_{ik}
(\eta'')\delta(\nu''-\nu_{ik})d\eta''=\left\{
\begin{array}{l}
u+\tau_{ik}(z_l),\, z\leq z_l,\\
u,\, z\geq z_l,\\
\end{array}
\right.
\label{int0}
\end{equation}
where
\begin{equation}
\tau_{ik}(z)=\frac{hc}{4\pi}\frac{n_i(z)B_{ik}}{H(z)}\left[1-\frac{n_k(z)g_i}
{n_i(z)g_k}\right],
\label{tauik}
\end{equation}
\begin{equation}
z_l=\frac{\nu_{ik}}{\nu}-1,\quad \frac{\nu_{ik}}{1+z_0}\leq \nu\leq\nu_{ik}.
\label{zl}
\end{equation}
When obtaining eq. (\ref{int0}) we turn to integration over $\nu''$ in the
middle part of this equation using relation
\begin{equation}
d\eta''=-\frac{c}{H(z'')}\frac{1+z''}{\nu''}d\nu'',
\end{equation}
which follows from equation $\nu''=(1+z'')\nu$ and eq. (\ref{etaz}). Here
$\nu$ is a radiation frequency in the present day epoch ($z=0$). So far as
the frequency does not change in the case of Thomson scattering and moreover
it does not change in fact in scattering in the lines in the adopted here
approximation then the frequency $\nu$ at $z=0$ is a parameter in the problem
and it can be not included into the arguments of the sought for solution.

Similarly one can transform the second (integral) term in the right side of
eq. (\ref{fsol}). As a result the formal solution (\ref{fsol}) becomes: 
$n(r,\mu,u)=n^0(r,\mu,u)$ for $u<u_l$ and
\begin{eqnarray}
&{\displaystyle n(r,\mu,u)=n_0(r_0,\mu_0)e^{-u-\tau_l}+
\int_{u_l}^u s(r',u')e^{u'-u}du'+}&\nonumber\\
&{\displaystyle +e^{-\tau_l}\int_0^{u_l}s^0(r',u')e^{u'-u}du'+
\lambda_l s(r'_l,u_l)e^{u_l-u}(1-e^{-\tau_l})}, \, u>u_l,&
\label{fsolu}
\end{eqnarray}
where $\tau_l=\tau_{ik}(z_l)$, $u_l=u(z_l)$, $\lambda_l=\lambda_{ik}(z_l)$,
$r'_l=r'|_{\eta'=\eta_l}$ and $n^0(r,\mu,u)$ and $s^0(r,u)$ are solutions
in the absence of scatterings in spectral lines (i.e. for $z_l=0$). By its
physical sence $z_l$ is a resonance redshift at which scattering of photons
with the frequency $\nu$ takes place and $\tau_l$ is the Sobolev optical
thickness of the medium for this redshift. The mentioned above quantities
appear in the theory of primordial hydrogen recombination lines formation
(see e.g. Dubrovich and Grachev, 2004).

By definition the source function $s(r,u)$ is an averaged over angles intensity
$j(r,u)$. Inserting eq. (\ref{fsolu}) into (\ref{jjik}) gives $s(r,u)=s^0(r,u)$
for $u<u_l$ while for $u>u_l$ it leads to the following integral equation
for $s(r,u)$:
\begin{eqnarray}
&{\displaystyle s(r,u)=s_0(r,u)e^{-u-\tau_l}+
\frac{1}{2r}\int_{u_l}^u e^{u'-u}
\frac{du'}{\eta-\eta'}\int_{|r-\eta+\eta'|}^{r+\eta-\eta'}
s(r',u')r'dr'+}& \nonumber\\
&{\displaystyle +\frac{e^{-\tau_l}}{2r}\int_0^{u_l} e^{u'-u}
\frac{du'}{\eta-\eta'}
\int_{|r-\eta+\eta'|}^{r+\eta-\eta'}s^0(r',u')r'dr'+}&\nonumber\\
&{\displaystyle+\frac{\lambda_l}{2r}(1-e^{-\tau_l})\frac{e^{u_l-u}}{\eta-\eta_l}
\int_{|r-\eta+\eta_l|}^{r+\eta-\eta_l}s^0(r',u_l)r'dr'
},&
\label{intequ}
\end{eqnarray}
where
\begin{equation}
s_0(r,u)=\frac{1}{2r\eta}
\int_{|r-\eta|}^{r+\eta}n_0(r_0,\mu_0)
r_0dr_0,
\label{s0intu}
\end{equation}
and according to eq. (\ref{rhomu0})
\begin{equation}
\mu=(r^2-r_0^2+\eta^2)/2r\eta,\quad
\mu_0=(r^2-r_0^2-\eta^2)/2r_0\eta.
\label{mumu0}
\end{equation}

When deriving the main eq. (\ref{intequ}) we turn from integration over $\mu$
to integration over $r'$ in the integral terms and to integration over $r_0$
in the free term. In the first case we use eq. (\ref{rhomup}) which yields
\begin{equation}
d\mu=-r'dr'/r(\eta-\eta'),
\end{equation}
and in the second case we use the first of eqs. (\ref{mumu0}) which gives
\begin{equation}
d\mu=-r_0dr_0/r\eta.
\end{equation}

It's worse noting that for $\tau_l=0$ the main eqs. (\ref{fsolu}) and
(\ref{intequ}) turn to the scalar equations obtained by us earlier (Grachev and
Dubrovich, 2011) for the case of purely Thomson scattring. As concerned to
the quantity of $\tau_l$ it does not exceed $4\cdot 10^{-4}$ according to
our calculations of primordial hydrogen recombination dynamics. So
it is expedient to expand solution in a power series of $\tau_l$ and to find the
first correction to the known solution for  $\tau_l=0$. So we seek solution in
the form
$s(r,u)=s^0(r,u)$ for $u<u_l$ and
\begin{equation}
s(r,u)=s^0(r,u)+\tau_l s^1(r,u),\, u>u_l.
\label{staul}
\end{equation}
Inserting this expression into the formal solution (\ref{fsolu}) and retaining
the terms not above the first power on $\tau_l$ leads to
\begin{equation}
\frac{n-n^0}{n^0}=-\tau_l\left\{1-\frac{\lambda_l}{n^0}\left[
e^{u_l-u}s^0(r'_l,u_l)+\int_{u_l}^u \tilde{s}(r',u')e^{u'-u}du'
\right]\right\},
\label{nexp}
\end{equation}
where $\tilde{s}=s^0+s^1$. Substitution of eq. (\ref{staul}) into eq.
(\ref{intequ}) gives for $\tilde{s}$ the following equation
\begin{equation}
\tilde{s}(r,u)=\tilde{s}_0(r,u)+\frac{1}{2r}\int_{u_l}^u e^{u'-u}
\frac{du'}{\eta-\eta'}\int_{|r-\eta+\eta'|}
^{r+\eta-\eta'}\tilde{s}(r',u')r'dr',\,u>u_l,
\label{tilds}
\end{equation}
where the free term
\begin{equation}
\tilde{s}_0(r,u)=\frac{1}{2r} \frac{e^{u_l-u}}{\eta-\eta_l}
\int_{|r-\eta+\eta_l|}^{r+\eta-\eta_l}s^0(r',u_l)r'dr'
\label{tilds0}
\end{equation}
is expressed through the source function $s^0(r,u)$ which is the solution of
the problem without scattering in spectral lines. Hence at first the function
$s^0(r,u)$ should be calculated by the same method as in our previous work
(Grachev and Dubrovich, 2011). Then eq. (\ref{tilds}) is solved and further
the dimensionless profile (\ref{nexp}) is calculated. In general case all
calculations must be fulfilled for each frequency separately. However in the
case of Thomson scattering radiation spectrum changes due to cosmological
redshift only. So if a frequency dependence of initial (for $u=0\,(z=z_0)$)
intensity $n_0(r,\mu,\nu)$ is detached from its spatial and angular dependences
i.e. if $n_0(r_0,\mu_0,\nu_0)=f(\nu_0)n_0(r_0,\mu_0)$ then using eq.
(\ref{nunu}) for $\nu_0$ the searched radiation intensity and source function
can be written as
\begin{equation}
n^0(r,\mu,u,\nu)=f(\nu(1+z_0)/(1+z))n^0(r,\mu,u),
\label{n0fn0}
\end{equation}
\begin{equation}
s^0(r,u,\nu)=f(\nu(1+z_0)/(1+z))s^0(r,u),
\label{s0fs0}
\end{equation}
where along the photon path $\nu(z)/(1+z)=\nu(0)$ is the photon frequency
in the present epoch (at $z=0$) so that the right sides (and hence the left
sides) of equations depend on the contemporaneous frequency. Substitution of
eqs. (\ref{n0fn0}) and (\ref{s0fs0}) into eq. (\ref{maineq}) (without an 
account of scattering in lines) instead of $n$ and $s$ leads to disappearance
of the derivative with respect to a frequency and after division of the both
sides of the equation by the constant multiplier $f(\nu(0)(1+z_0))$ we obtain
an equation for intensity $n^0(r,\mu,u)$ independent on frequency. We solved
such an equation earlier (Grachev and Dubrovich, 2011) by reducing it to an
integral equation for a source function $s^0(r,u)$. Further we used obtained
solutions to solve eq. (\ref{tilds}). It turned out that the function
$\tilde{s}(r,u)$ is very close to $s^0(r,u)$ and eq. (\ref{nexp}) can be
written as
\begin{equation}
\frac{n-n^0}{n^0}=-\tau_l\left\{1-\frac{\lambda_l}{n^0}\left[
e^{u_l-u}s^0(r'_l,u_l)+\int_{u_l}^u s^0(r',u')e^{u'-u}du'
\right]\right\}.
\label{nexp0}
\end{equation}
Here the right side does not depend on an initial radiation spectrum because it
enter in $n^0$ and $s^0$ as a constant multiplier $f(\nu(0)(1+z_0))$. By
designation of the right side of eq. (\ref{nexp0}) as $\varphi_{\nu}(r,\mu,u)$
where $\nu\equiv \nu(0)$ is a contemporaneous frequency we obtain that
according to eq. (\ref{nexp0}) radiation intensity (in terms of the mean
photons occupation numbers) is defined as follows
\begin{equation}
n(r,\mu,u_0,\nu)=f(\nu(1+z_0))n^0(r,\mu,u_0)[1+\varphi_{\nu}(r,\mu,u_0)],
\label{nrmun}
\end{equation}
where $\nu$ is a frequency, $\arccos\,\mu$ is an angular distance from the
burst center, $r$ is a distance from the burst center in the present day epoch
(for $z=0$($u=u_0))$. Here both dimensionless profile $\varphi_\nu$ and $n^0$
do not depend on a spectrum $f(\nu_0)$ of initial radiation.

As concerns $n^0$ we assume that at the initial time moment $t=0$ ($\eta=0$)
corresponding to some redshift $z_0$ it does not depend on angular variable
$\mu$ and it has spherically-symmetrical distribution:
\begin{equation}
n^0(r,\mu,0)=n_0(r),
\end{equation}
where $n_0(r)$ is defined as
\begin{equation}
n_0(r)=\pi^{-3/2}r_*^{-3}\exp[-(r/r_*)^2]\rightarrow \delta(r)/
4\pi\,r^2\quad \mbox{при}\quad r_*\rightarrow 0,
\label{n0rho}
\end{equation}
where $r_*$ is a parameter which defines characteristic size of the burst.

\begin{center}
{\bf Method of solution and main results}
\end{center}

Using the codes devised in our previous work (Grachev and Dubrovich, 2011) we
build summary profiles of H$_\alpha$ and H$_\beta$ lines and (separately) of
P$_\alpha$ and P$_\beta$ lines. The calculations were carried out for two
initial time moments $z_0$: 1600 and 1200 and for a few angular distances from
the direction to the burst center.

For the width of initial intensity distribution as a functin of $r$ (see eq.
(\ref{n0rho})) we use two values $r_*=1.5$ and 50 Mpc in a distance scale at
$z=0$. But in a distance scale corresponding to a burst moment (at $z=z_0$), 
the width of initial distribution for $z_0\gg 1$ will be significantly lower:
$a(z_0)r_*=r_*/(1+z_0)$.

As to another parameters in the problem they enter in particular in the
Hubble factor
\begin{equation}
H(z)=H_0\sqrt{\Omega_{\Lambda}+(1-\Omega)(1+z)^2+\Omega_{M}(1+z)^3+\Omega_{rel}
(1+z)^4},
\end{equation}
where $H_0=2.4306\cdot 10^{-18}h_0$ s$^{-1}$, $h_0$ is the Hubble constant in
units 75 km/(s$\cdot$Mpc); $\Omega_{M}$, $\Omega_{\Lambda}$ and $\Omega_{rel}$
are the density ratios of the matter, dark energy and relativistic particles
(radiation, massless neutrino) to the critical density $\rho_c=3H_0^2/(8\pi 
G)$ in the present epoch; $\Omega=\Omega_{M}+\Omega_{\Lambda}+\Omega_{rel}$,
$\Omega_{rel}=\rho^0_R(1+f_n)/\rho_c$, $\rho^0_R=a_RT_0^4/c^2$ is the density
of radiation mass now ($T_0$ is an average CMB temperature), $f_n$ is the part
of relativistic (massless) neutrino (usually $f_n=0.68$). For the flat model
of the Universe $\Omega=1$ and then $\Omega_{M}=1-\Omega_{\Lambda}-
\Omega_{rel}$.

Furthermore an electron concentration enters into equations. Usually it is
measured in the units of the total number density $n_{\rm H}$ of hydrogen
atoms and ions: $n_e(z)=x_e(z)n_{\rm H}(z)$ where $x_e(z)$ is so-called
recombination history of the Universe and
\begin{equation}
n_{\rm H}(z)=n_{\rm H}^0(1+z)^3,\quad  n_{\rm H}^0=0.63144\cdot10^{-5}X
\Omega_{\rm B}h_0^2 \,\mbox{cm}^{-3},
\end{equation}
where $\Omega_{B}$ is the ratio of the baryons density to the crytical density
now, $X$ is the hydrogen mass abundance. Recombination history is calculated
separately and it is an entry file. We calculate it using the code recfast.for
(Seager et al., 1999). As the base values we use: $\Omega=1$,
$\Omega_\Lambda=0.7$, $\Omega_{\rm B}=0.04$, $T_0=2.728$ K, hydrogen abundance
$X=0.76$, $\Omega_{rel}=0.85\cdot 10^{-4}$, Hubble constant $H_0=70$
km/(s$\cdot$Mpc).

As to the optical depths $\tau_{ik}$ in subordinate lines they were calculated
by means of our codes (Dubrovich and Grachev, 2004) for the learning of
primordial hydrogen recombination and occupation numbers of levels behavior.
Results of these calculations were also used to obtain time dependencies
of single scattering albedos $\lambda_{ik}(z)$.

On the fig. 1 are the graphs of the optical depths in H$_\alpha$, H$_\beta$,
P$_\alpha$ and P$_\beta$ lines and on the fig. 2 are the graphs of albedos
of single scattering in these lines. On the fig. 3 are the space profiles
of the average intensity of scattered radiation in the present epoch ($z=0$).

The results of the profiles calculations for H$_\alpha$ and H$_\beta$ lines are
presented on figs. 4 -- 8. The typical feature of the profiles is the presence
of absorption jumps caused by the source (burst) appearance at a given
redshift $z_0$ so that jumps arise on the frequencies $\nu=\nu_{ik}/(1+z_0)$
where $\nu_{ik}$ is the frequency of transition $i\rightarrow k$. Another
feature consists in the presence of emission components which appear in the
profiles for distances on the back side of spatial distribution of average
radiation intensity (see fig. 3) where photons ``lagged behind'' due to
scatterings are disposed. Photons scattered in line sideways return into a given
direction due to scatterings on free electrons with some delay. Just they
form emission components of the line profiles and there is a minimum of
``lagged behind'' photons in the direction of the burst center.
So when moving off this direction emission components enhance and absorption
components weaken (see fig. 8). When passing from the back front of an average
intesity distribution to the front one (see crosses on the curves on fig. 3)
emission components weaken and absorption ones enhance and in the end the line
becomes purely absorption one (see figs. 4 -- 7). This is connected with the
decrease of contribution of photons ``lagged behind'' due to scatterings.

From comparison of the profiles for the bursts with great (50 Mpc) and small
(1.5 Mpc) size it follows that in the first case absorption is less deep than
in the second one. Further an appearance of narrow emission component on the
fig. 7 for the burst at $z_0=1200$ with characteristic size 1.5 Mpc is
connected with a comparatively small optical thickness ($\approx 3$) of the
Universe due to Thomson scattering between the moments $z=1200$ and $z=0$.
So we see photons arrived from the burst directly without scatterings.
On the fig. 3 they form a narrow peak with a width $\approx 1.5$ Mpc at the
distance equal to conformal time of the burst center $\eta=13635$ Mpc. On the
fig. 7, where the profiles for different distances from the burst center are
displayd, the maximum height of emission peak is achieved namely on the
distance close to mentioned above conformal time. On the smaller distances
overwhelming contribution give scattered photons which have significantly
wider density redistribution in space (see fig. 3). So the height of emission
peaks turns out to be smaller and their width greater.

Also the profiles for the case of pure scattering ($\lambda_{23}=\lambda_{24}=
1$) were calculated and they were compared with the profiles for real
dependence of a single scattering albedo on redshift $z$ (see fig. 2). As it
should be expected in the last case the jumps of absorption appear on the
frequencies corresponding to the moments of the bursts appearance and as a
whole the profiles go lower than in the case of pure scattering. The
difference is especially great for the burst of large size.

The summary profiles of P$_\alpha$ and P$_\beta$ lines were also calculated.
P$_\alpha$ and P$_\beta$ lines turn out to be 10--30 times weaker than
H$_\alpha$ and H$_\beta$ lines because of the lower optical depths (see fig. 1)
and moreover they are much more weaker depend on parameters because of the lower
albedo of single scattering (see fig. 2). In the limit of
$\lambda\rightarrow 0$ the line profile is purely absorption one and it is
defined only by an optical depth profile (as it is seen from eq. (\ref{nexp0}))
and by the burst moment $z_0$ which defines position of the jump in the profile
at the low frequency edge for $\nu=\nu_{ik}/(1+z_0)$.

It should be stressed once more that the profiles displayd on figs. 4 -- 7 are
calculated with the proper account of the real dependence of single scattering
albedos $\lambda_{ik}$ in lines on redshift $z$.

\begin{center}
{\bf Conclusions}
\end{center}

The local bursts model of primordial plasma and radiation temperature 
fluctuations in the early Universe is considered. These fluctuations can be
represented as fastly variable sources with the initial blackbody spectrum
of radiation with the temperature which slightly differs from the average
CMB temperature. More generally, one may consider as a source any real picked
out object: primordial accreting black hole for example. In this work we
calculate transitional function from radiation intensity of these sources to
an observed intensity taking into account photons scatterings on free
electrons and in hydrogen subordinate lines. In the first approximation this
function makes sense of an optical depth in regard to scattering in lines.
However due to multiplicity of scattering on free electrons a part of
photons scattered in atomic lines returns on the line of sight in another
space point which leads to an appearance of emission components in line
profiles.

In the capacity of the model we consider scattering of continuous radiation
of the source (instantaneous burst of radiation at a given redshift $z_0$) on
electrons and in H$_\alpha$, H$_\beta$, P$_\alpha$ and P$_\beta$ lines of
primordial hydrogen at the recombination epoch. It is shown that thus arising
lines in a source spectrum are generally absorption ones with the depths from
$3\cdot 10^{-5}$ to $2\cdot 10^{-4}$ (for H$_\alpha$ line) depending on
characteristic initial size of the source and on the distance from its
center. For P$_\alpha$ line the depth of absorption is by 10 -- 30 times
smaller. Real observations give maximum amplitude of the order of 500 $\mu$K for
temperature deviation in the spots. So relative magnitude of temperature
fluctuations lies within the limits of 10$^{-7}$ -- 10$^{-9}$. The profiles
may contain emission components with lower intensities (with respect to
source continuum) as compared with absorption components. It's worse noting
also that the most deep absorption is at the low frequency edge of the profile.

This work was supported in part by St. Petersburg State University grant No.
6.38.18.2014.
\newpage

\begin{center}
{\bf References}
\end{center}

\begin{enumerate}
\item  Adam R., Ade P.A.R., et al., Planck 2015 results. I. Overview
of products and scientific results, Planck collaboration, arXiv: 1502.01582v1
[astro-ph.CO] (2015a).
\item  Adam R., Ade P.A.R., et al., Planck 2015 results. IX. Diffuse
component separation: CMB maps, Planck collaboration, arXiv: 1502.05956v1
[astro-ph.CO] (2015b).
\item  Grachev S.I., Dubrovich V.K., Astrophysics {\bf 34}, 124 (1991).
\item  Grachev S.I., Dubrovich V.K., Astron. Lett. {\bf 37}, 293 (2011).
\item  Dubrovich V.K., Astron. Lett. {\bf 29}, 6 (2003).
\item  Dubrovich V.K., Grachev S.I., Astron. Lett. {\bf 30}, 657 (2004).
\item  Dubrovich V.K., Glazyrin S.I., Cosmological dinosaurs,
arXiv:1208.3999v1 [astrp-ph.CO] 20 Aug (2012).
\item  Nagirner D.I., Kirusheva S.L., Astron. Rep. {\bf 49}, 167 (2005).
\item  Rubi{\~n}o-Martin J.A., Hern{\'a}ndez-Monteagudo C.,
Sunyaev R.A., Astron. Astrophys., {\bf 438}, 461 (2005).
\item  Seager S., Sasselov D.D., Scott D., Astrophys. J. Suppl. 
Ser. {\bf 128}, 407 (2000). 
\end{enumerate}

\newpage
\begin{figure}[p]
\centering

\resizebox{1.0\textwidth}{!}{\includegraphics{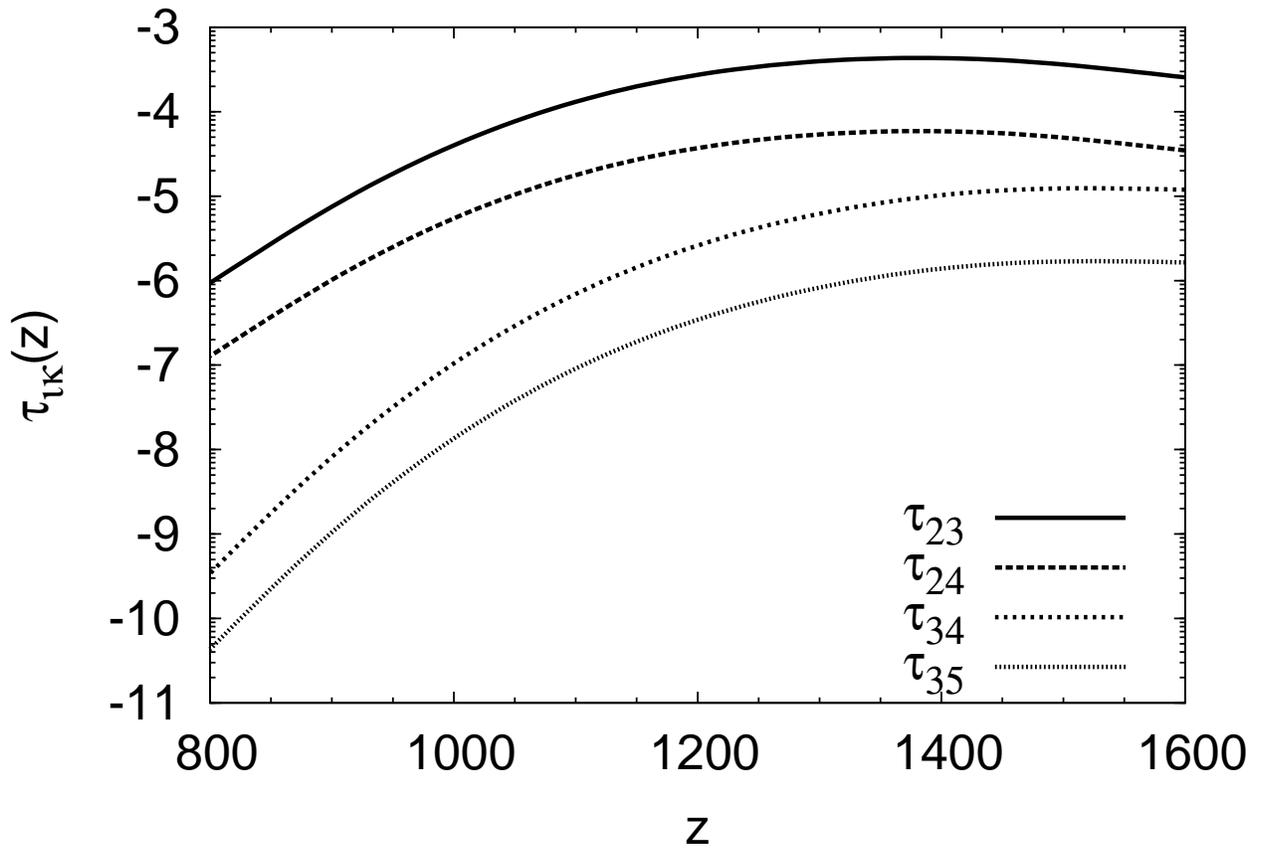}}

\caption{The profiles of optical depths in H$_\alpha$, H$_\beta$,
P$_\alpha$ and P$_\beta$ lines.}

\end{figure}
\begin{figure}[p]
\centering

\resizebox{1.0\textwidth}{!}{\includegraphics{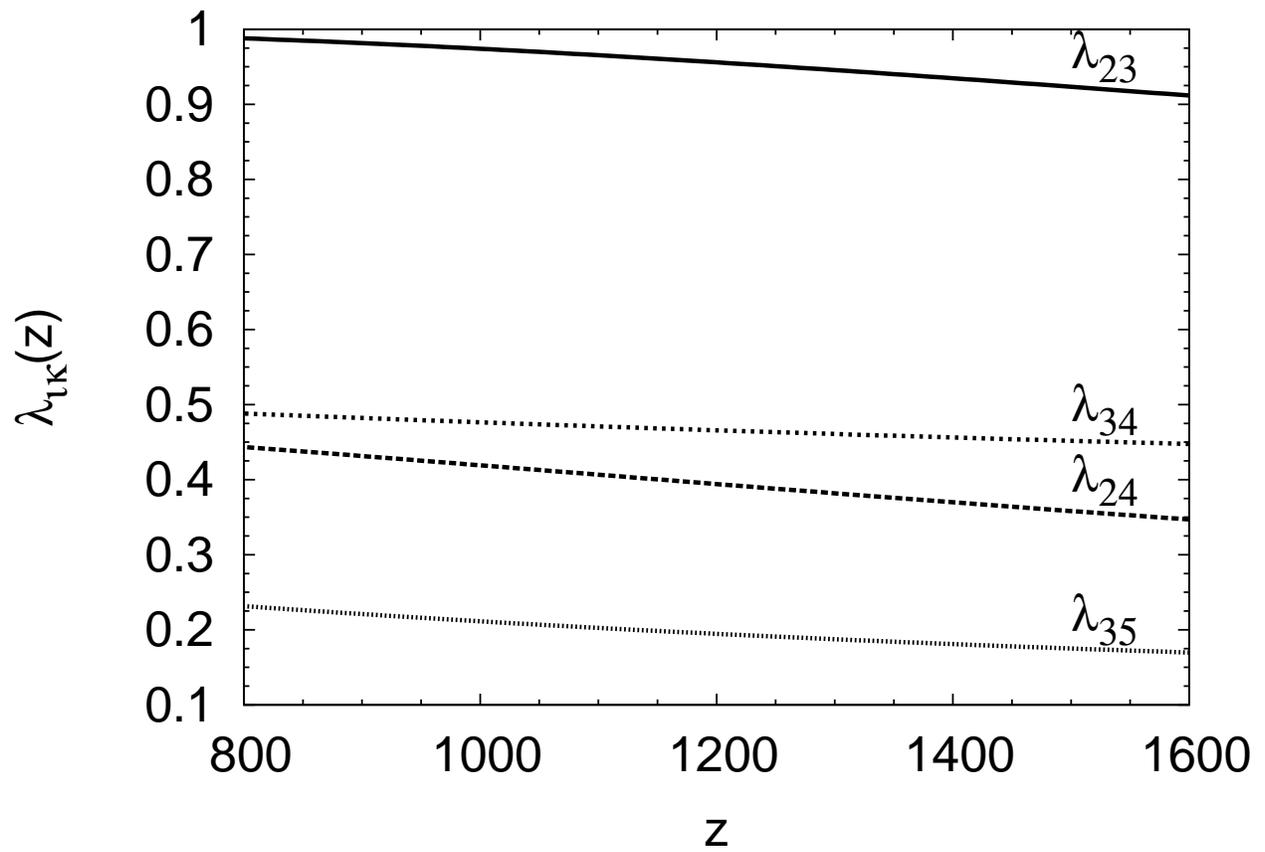}}

\caption{The profiles of albedos of a single scattering in H$_\alpha$,
H$_\beta$, P$_\alpha$ and P$_\beta$ lines.}

\end{figure}
\begin{figure}[p]
\centering

\resizebox{1.0\textwidth}{!}{\includegraphics{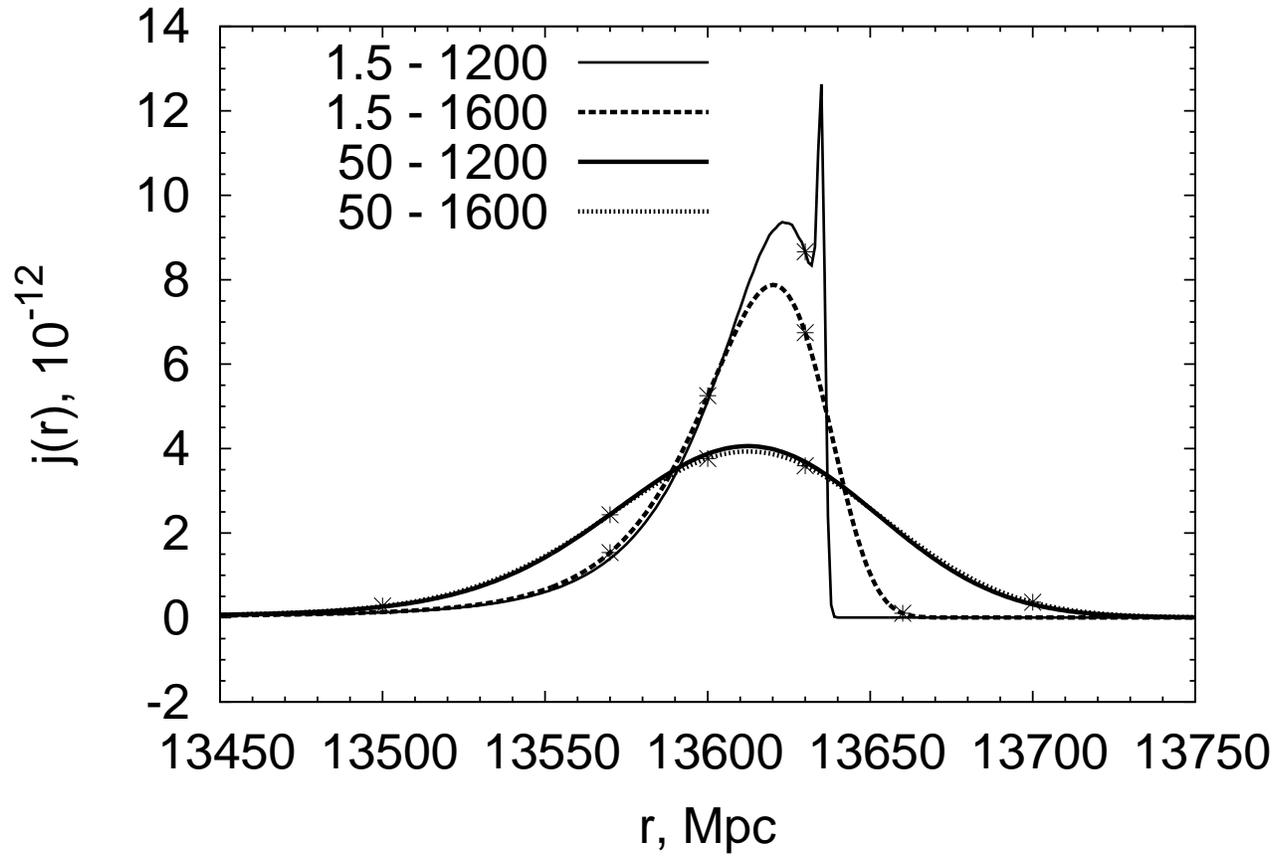}}

\caption{The profiles of average intensity of scattered radiation for the
bursts at $z_0=1200$ and 1600 with characteristic sizes 1.5 and 50 Mpc
(at $z=0$). The crosses mark the points where the profiles of subordinate
lines were calculated.}

\end{figure}

\begin{figure}[p]
\centering

\resizebox{1.0\textwidth}{!}{\includegraphics{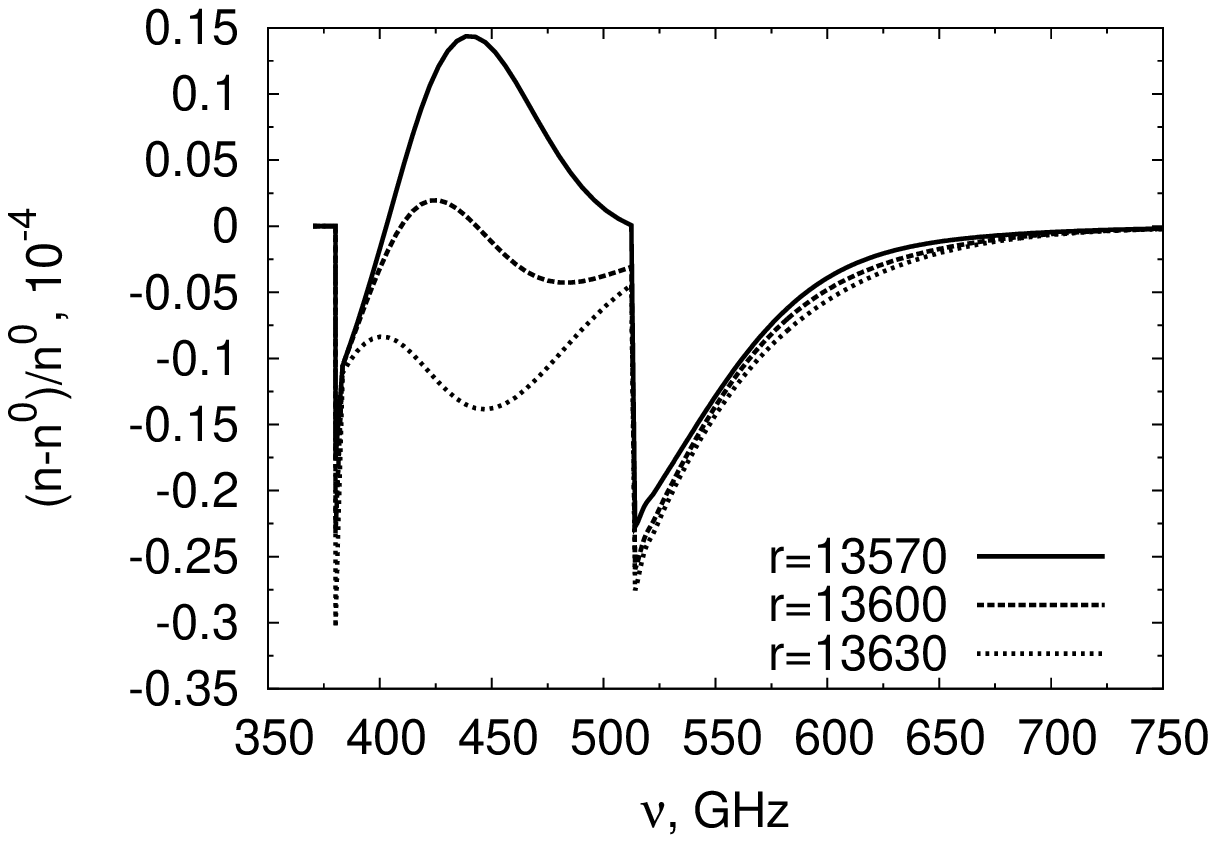}}

\caption{Summary profiles of H$_\alpha$ and H$_\beta$ lines for the burst
at $z_0=1200$ with the characteristic size 50 Mpc (at $z=0$) in the direction
of the burst center for different distances $r$ (in Mpc) from the center.}

\end{figure}

\begin{figure}[p]

\centering

\resizebox{1.0\textwidth}{!}{\includegraphics{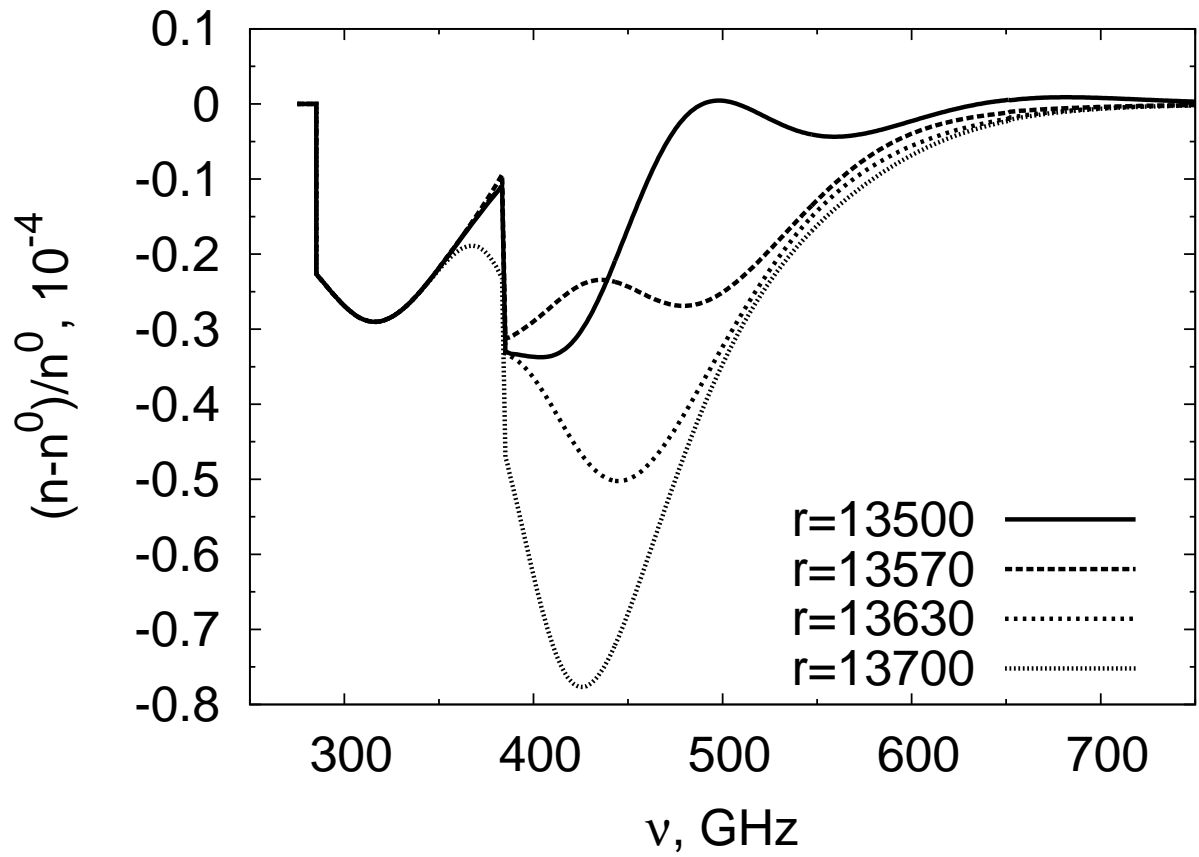}}

\caption {The same as on fig. 4 but for $z_0=$1600.}

\end{figure}

\begin{figure}[p]
\centering

\resizebox{1.0\textwidth}{!}{\includegraphics{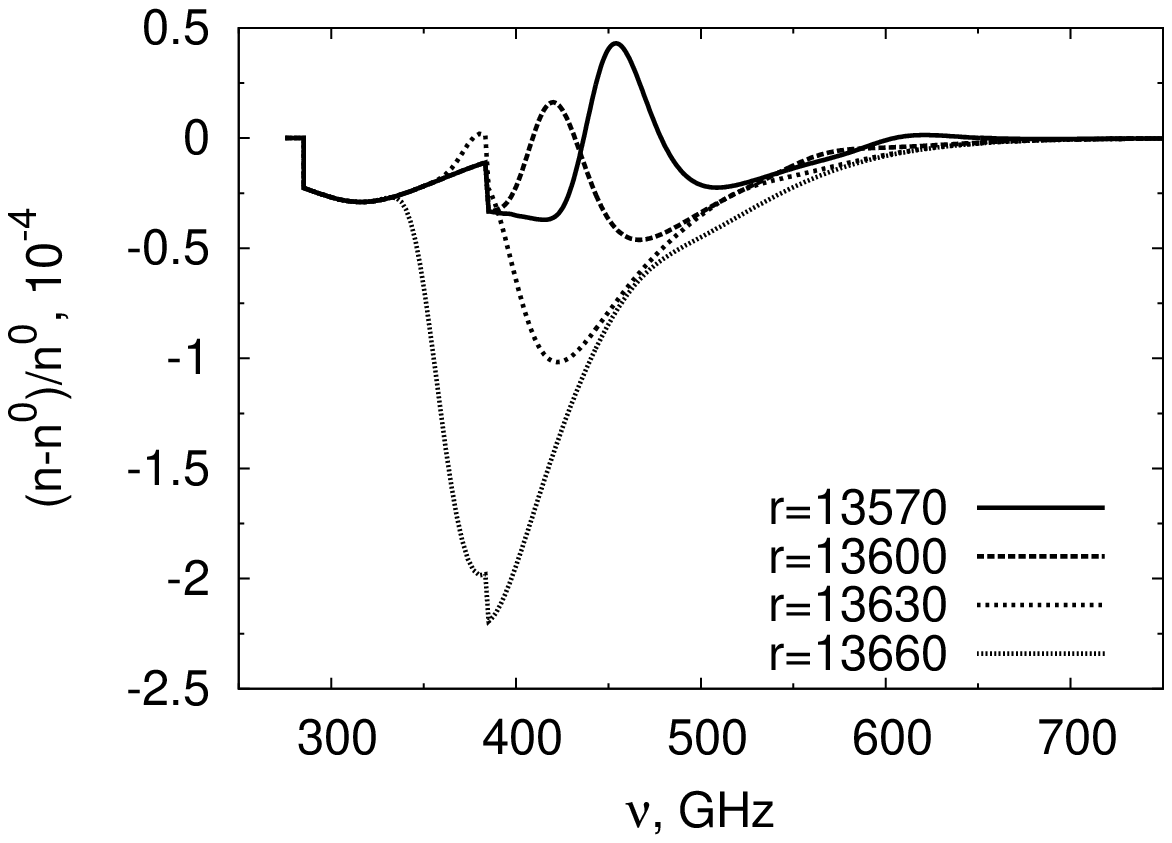}}

\caption{Summary profiles of H$_\alpha$ and H$_\beta$ lines for the burst
at $z_0=1600$ with the characteristic size 1.5 Mpc (at $z=0$) in the
direction of the burst center for different distances $r$ (in Mpc) from the
center.}

\end{figure}

\begin{figure}[p]

\centering

\resizebox{1.0\textwidth}{!}{\includegraphics{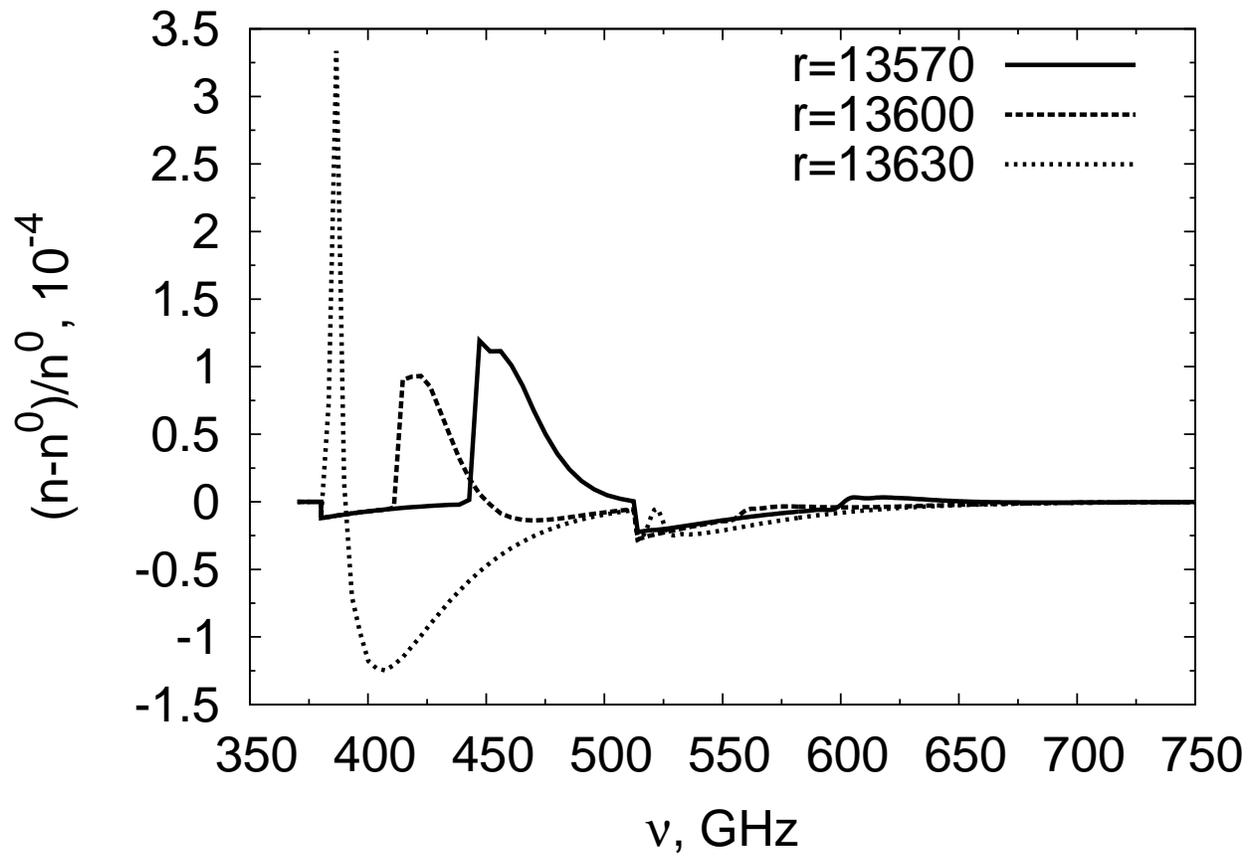}}
\caption
{The same as in the preceeding figure but for z$_0=1200$.}
\end{figure}

\begin{figure}[p]

\centering
\resizebox{1.0\textwidth}{!}{\includegraphics{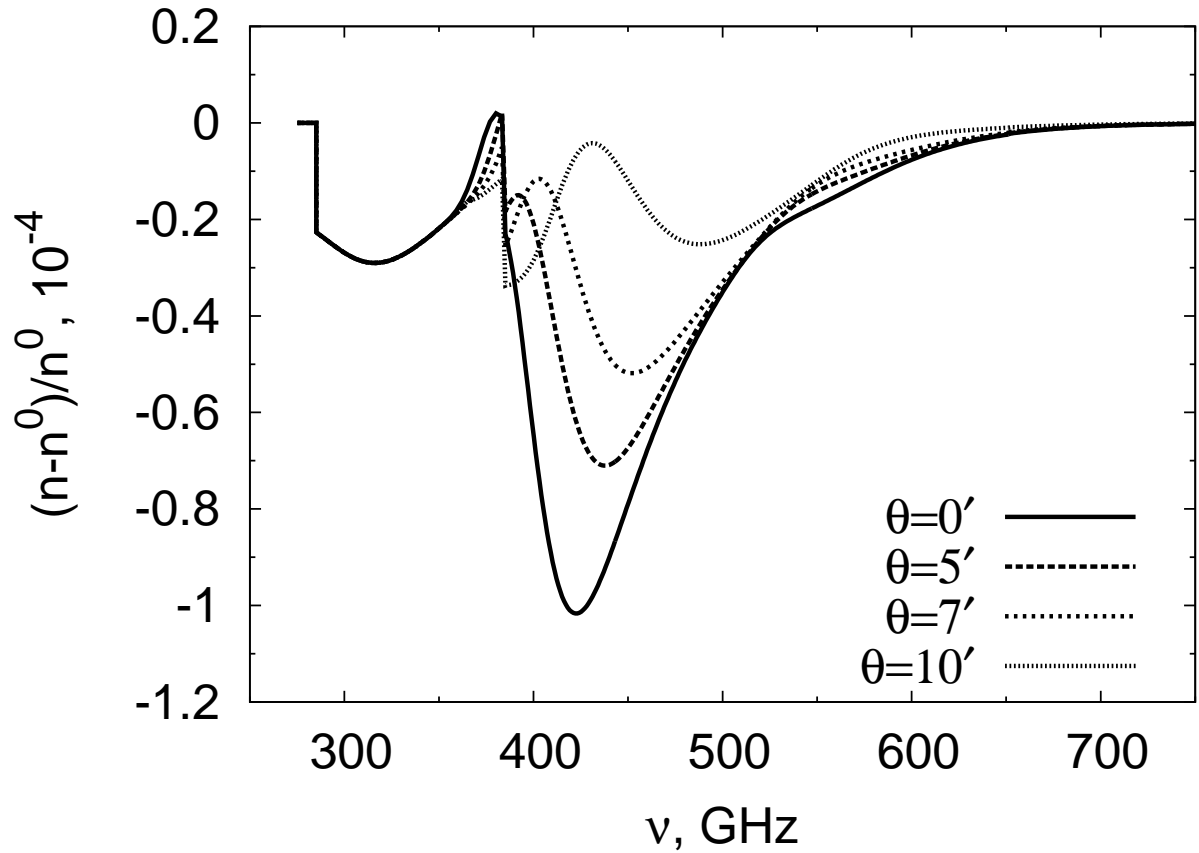}}
\caption{Summary profiles of H$_\alpha$ and H$_\beta$ lines on different
angular distances $\theta$ from the burst center for the burst at $z_0=1600$
with the characteristic size $r_*=1.5$ Mpc. The distance from the center of
the burst is $r=13630$ Mpc.}

\end{figure}

\newpage

\end{document}